\documentclass[useAMS,usenatbib]{mn2e}
\usepackage{txfonts}
\usepackage{times}
\usepackage{graphicx}% Include figure files
\usepackage{dcolumn}% Align table columns on decimal point
\usepackage{bm}% bold math
\usepackage{epsfig}
\usepackage{color}

\def\fun#1#2{\lower3.6pt\vbox{\baselineskip0pt\lineskip.9pt
  \ialign{$\mathsurround=0pt#1\hfil##\hfil$\crcr#2\crcr\sim\crcr}}}
\def\simgt{\mathrel{\lower0.6ex\hbox{$\buildrel {\textstyle >}
 \over {\scriptstyle \sim}$}}}
\def\simlt{\mathrel{\lower0.6ex\hbox{$\buildrel {\textstyle <}
 \over {\scriptstyle \sim}$}}}

\input epsf

\newcommand{\hompc}{\,h\,{\rm Mpc}^{-1}}
\newcommand{\mpcoh}{\,h^{-1}\,{\rm Mpc}}

\newcommand{\aj}{AJ}
\newcommand{\apj}{ApJ}

\newcommand{\mnras}{MNRAS}

\newcommand{\aap}{A{\&}A}

\newcommand{\prd}{Phys. Rev. D}
\newcommand{\apjl}{ApJL}

\newcommand{\nat}{Nature}

\def\be{\begin{equation}}
\def\ee{\end{equation}}
\def\ba{\begin{eqnarray}}
\def\ea{\end{eqnarray}}

\def\nn{\nonumber}

\begin{document}

%\preprint{}

\title{Decomposition of Spectra from Redshift Distortion Maps} 
 
\author[Song and Kayo]{
\parbox{\textwidth}{
Yong-Seon Song$^{1}$ and Issha Kayo$^{2}$}
\vspace*{4pt} \\
$^{1}$ Institute of Cosmology $\&$ Gravitation, University of Portsmouth, Portsmouth, PO1 3FX, UK\\
$^{2}$ Institute for the Physics and Mathematics of the Universe, University of Tokyo, 5-1-5 Kashiwanoha, Chiba 277-8582, Japan}

\date{\today} 
\pagerange{\pageref{firstpage}--\pageref{lastpage}}

\maketitle

\begin{abstract}
We develop an optimized technique to extract density--density and velocity--velocity spectra out of observed spectra in redshift space. The measured spectra of the distribution of halos from redshift distorted mock map are binned into 2--dimensional coordinates in Fourier space so as to be decomposed into both spectra using angular projection dependence. With the threshold limit introduced to minimize nonlinear suppression, the decomposed velocity--velocity spectra are reasonably well measured up to scale $k=0.07\hompc$, and the measured variances using our method are consistent with errors predicted from a Fisher matrix analysis. The detectability is extendable to $k\sim 0.1\hompc$ with more conservative bounds at the cost of weakened constraint.
\end{abstract}

%\pacs{draft}

%\keywords{Large-scale structure formation}

\section{Introduction}

The evolution of large scale structure, as revealed in the clustering of galaxies observed in wide--deep redshift surveys has been one of key cosmological probes. Structure formation is driven by a competition between gravitational attraction and the expansion of space-time, which enables us to test our model of gravity at cosmological scales and the expansion of history of the Universe~\citep{Wang:2007ht,Linder:2007nu,Guzzo:2008ac,2009JCAP...10..004S,Simpson:2009zj,Guzik:2009cm,McDonald:2008sh,Stril:2009ey,Bean:2010zq}. 

Maps of galaxies where distances have been measured from redshifts show
anisotropic deviations from the true galaxy distribution
\citep{2000AJ....120.1579Y,2001Natur.410..169P,2003astro.ph..6581C,2003MNRAS.346...78H,2004MNRAS.353.1201P,2005ApJ...630....1Z,2005AA...439..877L,2006PhRvD..74l3507T,2008ApJ...676..889O,2008arXiv0812.2480G,2008AA...486..683G,Guzzo:2008ac},
because galaxy recession velocities include components from both the
Hubble flow and peculiar velocities. In linear theory, a distant
observer should expect a multiplicative enhancement of the overdensity
field of tracers due to the peculiar motion along the line of sight \citep{Davis:1982gc,1987MNRAS.227....1K,1989MNRAS.236..851L,1990MNRAS.242..428M,1991MNRAS.251..128L,1992ApJ...385L...5H,1994MNRAS.266..219F,1995ApJ...448..494F}. In principle, the observed spectra in redshift space can be decomposed into both density--density and velocity--velocity spectra using angular projection dependence~\citep{2009JCAP...10..004S,Percival:2008sh,2009MNRAS.397.1348W,Song:2010vh}. With a local linear bias, the real-space galaxy density field is affected, while the peculiar velocity term is not. In this paper, we attempt to extract velocity--velocity spectra as an unbiased tool to trace the history of structure formation.

A theoretical formalism~\citep{2009MNRAS.397.1348W} was derived for forecasting errors when extracting velocity--velocity spectra out of the observed redshift space distortion maps. However, it is not yet fully understood what the optimal technique is to practically decompose the spectra as theory predicts.
We propose a statistical technique to extract it up to the limit of
theoretical estimation. Our method utilizes the distinct angular
dependence of density--density and velocity--veclocity spectra to
decompose them from two--dimensional redshift power spectra,
and is consistent with the theoretical estimate from Fisher matrix analysis.

We present the detailed formalism in the next section. The Fisher matrix
analysis to decompose spectra is briefly reviewed, then we present the 
method to decompose spectra in an optimal way with mock data. We discuss
statistical method to minimize the effect by nonlinear suppression. 

\section{Peculiar velocity power spectra extraction}

\subsection{Theoretical Expectation of Decomposition Accuracy}

The observed power spectrum in redshift space is decomposed into spectra
of density fluctuations and peculiar velocity fields in real space. The
observed power spectra in redshift space, $\tilde{P}$, is given by,
\begin{eqnarray}
\tilde{P}(k,\mu,z) &=& \big\{P_{gg}(k,z)
   + 2\mu^2r(k)\left[P_{gg}(k,z)P_{\Theta\Theta}(k,z)\right]^{1/2}\nonumber\\
  &+& \mu^4P_{\Theta\Theta}(k,z)\big\}G(k,\mu,\sigma_v),
\end{eqnarray}
where
$P_{gg}$ is the galaxy--galaxy density spectrum, $P_{\Theta\Theta}$ is the velocity--velocity
spectrum ($\Theta$ is the divergence of velocity map in unit of $aH$), and
$\mu$ denotes the cosine of the angle between orientation of the wave
vector and the line of sight. Because this decomposition is valid only
at large scale and when the rotation of the velocity field is negligible, we
focus on modes of $k<0.1\hompc$ \citep{Pueblas:2009qy}. The
cross-correlation coefficient $r(k)$ is defined as $r(k)\equiv
P_{g\Theta}/\sqrt{P_{gg}P_{\Theta\Theta}}$. The density and velocity
divergence are highly correlated for $k<0.1\hompc$ so we assume that
both are perfectly correlated, $r(k)\sim
1$~\citep{2009MNRAS.397.1348W}. Then the density-velocity cross-spectrum
becomes the geometric mean of the two auto-spectra and we have only two
free functions, $P_{gg}$ and $P_{\Theta\Theta}$. As \cite{Scoccimarro04}
clearly pointed out, the redshift space power spectrum is suppressed
along line-of-sight due to the velocity dispersion of large-scale flow,
and we follow his model by introducing a function
$G=\exp(-k^2\mu^2\sigma_v^2)$ where $\sigma_v$ will be calculated from
linear theory. Considering the possibility that nonlinear
dynamics, like Finger-of-Gods effect, might contaminate the power
spectrum, we use this term to
find a cut-off scale of $\mu$ to exclude data which could be affected
strongly by nonlinear dynamics. Indeed, \cite{2009PhRvD..80l3503T}
pointed out that $\sigma_v$ calculated by linear theory does not match
with result from N-body simulations if one tries to model the power
spectrum at $\gtrsim 0.1 \hompc$.
This cut-off edge $\mu_{\rm cut}$
is defined by $\mu_{\rm cut} \equiv \sigma_{\rm th}/k\sigma_v$, where
the value of $\sigma_{\rm th}$ will be discussed later.

\begin{figure}
\begin{center}
\resizebox{3.2in}{!}{\includegraphics{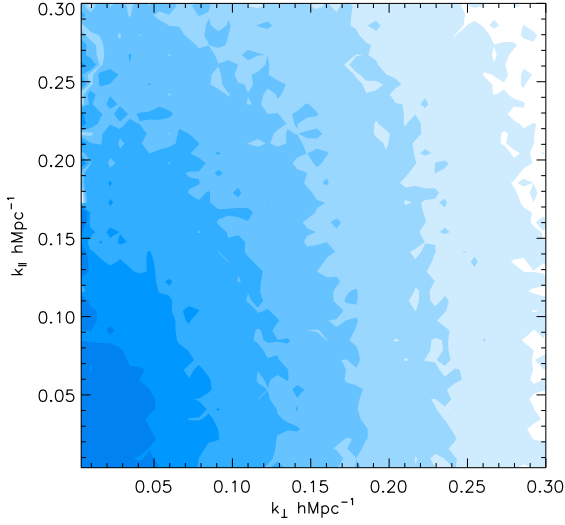}}
\end{center}
\caption{Power spectra from mock map in 2D cartesian coordinate $(k_{\bot}, k_{\|})$.}
\label{fig:contour_Pk}
\end{figure}

We estimate the accuracy of decomposition of $P_{gg}$ and $P_{\Theta\Theta}$ from $\tilde{P}$ using Fisher matrix analysis determining the sensitivity of a particular measurement. Fisher matrix for this decomposition, $F_{\alpha\beta}^{\rm dec}$, is written as,
\ba\label{eq:Fdec}
F_{\alpha\beta}^{\rm dec}=\int^{\mu_{\rm cut}}_{-\mu_{\rm cut}}d\mu\int\frac{\partial\tilde{P}(k,\mu)}{\partial p_{\alpha}}\frac{\partial\tilde{P}(k,\mu)}{\partial p_{\beta}}\frac{V_{\rm eff}(\tilde{P})}{\tilde{P}(k,\mu)^2}\frac{k^2dk}{2(2\pi)^2}\,,
\ea
where $p_{\alpha}=(P_{gg},P_{\Theta\Theta})$. The effective volume $V_{\rm eff}(\tilde{P})$ is given by,
\ba
V_{\rm eff}(\tilde{P})=\left[\frac{n\tilde{P}}{n\tilde{P}+1}\right]^2V_{\rm survey}\,,
\ea
where $n$ denotes galaxy number density. 

Derivative terms in Eq.~(\ref{eq:Fdec}) are given by,
\begin{eqnarray}
  \frac{\partial \ln \tilde{P}(k_i,\mu,z_j)}{\partial P_{gg}(k_i,z_j)}
  &=& \frac{1}{\tilde{P}(k_i,\mu,z_j)}
  \left[1 + \mu^2
  \sqrt{\frac{P_{\Theta\Theta}(k_i,z_j)}{P_{gg}(k_i,z_j)}} 
  \right] \nonumber \\
  \frac{\partial\ln \tilde{P}(k_i,z_j)}{\partial P_{\Theta\Theta}(k_i,z_j)}
  &=&\frac{\mu^2}{\tilde{P}(k_i,\mu,z_j)}
  \left[\sqrt{\frac{P_{gg}(k_i,z_j)}{P_{\Theta\Theta}(k_i,z_j)}}+\mu^2\right]\,.
\end{eqnarray}
The diagonal elements of the inverse Fisher matrix indicate the estimated errors of decomposion accuracy. The variances of $P_{gg}(k_i,z_j)$ and $P_{\Theta\Theta}(k_i,z_j)$ is given by,
\ba\label{eq:sigP}
\sigma[P_{gg}(k_i,z_j)] &=& \sqrt{F_{gg}^{\rm dec\,-1}(k_i,z_j)} \nn\\
\sigma[P_{\Theta\Theta}(k_i,z_j)] &=& \sqrt{F_{\Theta\Theta}^{\rm dec\,-1}(k_i,z_j)}\,.
\ea

\begin{figure}
\begin{center}
\resizebox{2.93in}{!}{\includegraphics{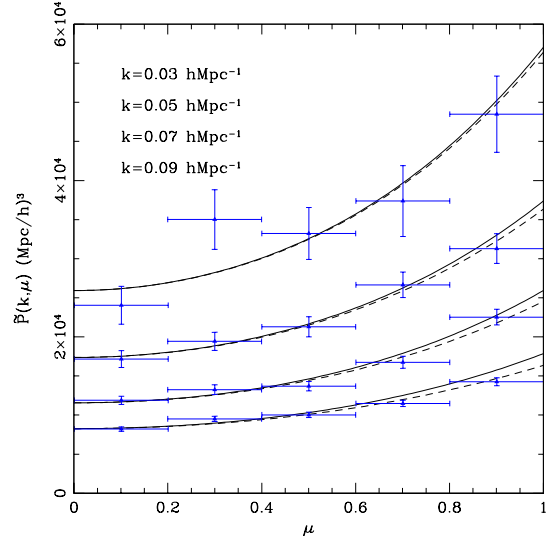}}
\end{center}
\caption{The observed power spectra at scales, $\bar{k}$=0.03, 0.05, 0.07 and 0.09$\hompc$ (from top to bottom) are plotted with error bars at various $\mu$. Solid curves are $\tilde{P}_{\rm th}(k,\mu)$ (Kaiser effect alone) and dash curves are $\tilde{P}_{\rm fit}(k,\mu)$ (including dispersion effect) from best fitting bias $b(k)$.}
\label{fig:Pks}
\end{figure}

\subsection{2D power spectra from mock map}

We use the halo catalogue from the time-streaming mock map of the Horizon simulation~\citep{Teyssier:2008zd},
and cut 1 (Gpc/$h$)$^3$ cubic box at the median redshift
$\bar{z}=0.83$, which contains 2.2 million halos. 
The fiducial cosmological parameters of the simulation
are given by $(\Omega_m=0.24, \Omega_k=0, h=0.72, \sigma_8=0.78,
n_S=0.96)$ and the initial transfer function is given by \cite{1998ApJ...496..605E}.

The distribution of halos is modified according to their peculiar
velocity to incorporate the redshift distortion effect. We adopt the
distant observer approximation and measure the power spectrum in
$(k_{\bot}, k_{\|})$ space. The density fluctuation field is constructed
by assigning the halos to $512^3$ grids for the fast Fourier
transformation (FFT) using the nearest grid point (NGP)
method. Fig.~\ref{fig:contour_Pk} shows the resulting power spectrum. While
linearly spaced bins in $(k_{\bot},k_{\|})$ are used in this plot for
presentation purpose, we use bins in $k$ and $\mu$ for the following
analysis. $k$ is divided in $\Delta k=0.02\hompc$ linearly equally spaced bins from $k=0.02\hompc$ to $0.2\hompc$ and $\mu$ is in 5 linear-bins from 0 to 1 with equal spacing. The measured 2D power spectra in ($k$,$\mu$) coordinate are shown in Fig.~\ref{fig:Pks}.

The Gaussian variance is used to derive errors for each bin shown as error bars in Fig.~\ref{fig:Pks}, $\sigma[\tilde{P}_{\rm ob}(k,\mu)]=\tilde{P}(k,\mu)\sqrt{2/N(k,\mu)}$ where $N(k,\mu)$ is number of modes in Fourier space. We test this using an alternative method, jack--knife errors (we do not attemp to generate more samples as we are interested in mocking real observables in a single patch). A total 64 jack--knife samples are prepared out of a single mock map by dividing each coordinate into 4 pieces. Both errors agrees well, and different bins weakly correlate with each other.

Halo distribution is a biased tracer of the dark matter distribution. Theoretical $\tilde{P}_{\rm th}(k,\mu)$ from Kaiser effect only is given by,
\ba
\tilde{P}_{\rm th}(k,\mu)=b^2P_{mm} + 2b\mu^2r_{\rm h}\sqrt{P_{mm}P_{\Theta\Theta}} + \mu^4P_{\Theta\Theta}\,,
\ea
where $P_{mm}(k)$ is the dark matter density--density spectra and $b=b(k)$ is
the halo bias for each given scale. Spectra $P_{mm}(k)$ and
$P_{\Theta\Theta}(k)$ are given from the cosmological parameters used for
the simulation, and the halo cross--correlation parameter $r_{\rm h}$ is set
to be unity. It has been tested that $r$ for dark matter--$\Theta$ is
perfectly correlated at linear scales $k<0.1\hompc$ from
simulation. Unfortunately, the same sanity check is not applicable for
halo maps due to the insufficient number of halo in each grid for direct
velocity power spectra. Instead, the theoretical $\tilde{P}_{\rm
th}(k,\mu)$ is derived based upon $r_{\rm h}(k)=1$, and the possible
departure from the unity is dectectable from measured
$P_{\Theta\Theta}(k)$ at linear scales.

\begin{table}
\begin{center}
\begin{tabular}{c|cccc}
 $k(\hompc)$ &$0.03$&$0.05$ &$0.07$ &$0.09$  \\
\hline
$b(k)$  &  $1.65^{\pm 0.44}$&  $1.70^{\pm 0.27}$   & $1.60^{\pm 0.18}$   & $1.69^{\pm 0.18}$ 
\end{tabular}
\end{center}
\caption{Best fitting biases $b(k)$ at given scales $k$ from $k=0.03$ to 0.09$\hompc$.}
\label{tab:bk}
\end{table}

The tracer bias is assumed not to be determined by theoretical formalism
or by other experiment. Instead of applying scale independent bias,
$b(k)$ is varied independently for each $k$--bin. We fit $b(k)$ for each
mode to get $\tilde{P}_{\rm th}(k,\mu)$ (solid curves in
Fig.~\ref{fig:Pks}). In Table~\ref{tab:bk}, the best fit $b(k)$ is given with 1--$\sigma$ confidence level. Theoretical $\tilde{P}_{\rm th}(k,\mu)$ with fitted $b(k)$ is over--plotted with the measured $\tilde{P}_{\rm ob}(k,\mu)$ from the simulation in Fig.~\ref{fig:Pks}. We cut out scales $k<0.03\hompc$ due to our limited box size and $k>0.1\hompc$ due to non-linear effects. 

Using $\tilde{P}_{\rm th}(k,\mu)$, theoretical errors are estimated from Fisher matrix analysis. Un-filled black contours in Fig.~\ref{fig:contour_dP_dec} represent the theoretical expectation around $b(k)^2P_{mm}(k)$ and $P_{\Theta\Theta}(k)$. As it is prediced from halo bias model, measured bias is nearly scale independent.

\subsection{Practical approach to extract peculiar velocity spectra}

\begin{figure}
\begin{center}
\resizebox{3.2in}{!}{\includegraphics{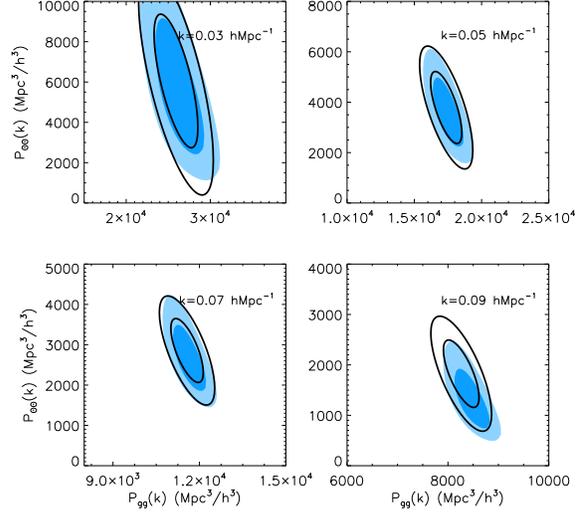}}
\end{center}
\caption{Contour plots are shown for decomposed $P_{gg}(k_i)$ and $P_{\Theta\Theta}(k_i)$ at $k=0.03$, 0.05, 0.07 and 0.09$\hompc$. Unfilled black contours represent theoretical prediction from Fisher matrix analysis, and filled blue contours represent measured $P_{gg}(k_i)$ and $P_{\Theta\Theta}(k_i)$ from mock map in redshift space.}
\label{fig:contour_dP_dec}
\end{figure}

Spectra $P_{gg}(k_i)$ and $P_{\Theta\Theta}(k_i)$ are fitted simultaneously to $\tilde{P}_{\rm ob}(k_i,\mu_p)$ where $i$ and $p$ denote $k$ and $\mu$ bins respectively. Bias is not parameterized to fit $\tilde{P}_{\rm ob}(k_i,\mu_p)$, instead, we use $P_{gg}(k_i)$. The fitting $\tilde{P}_{\rm fit}(k_i,\mu_p)$ is given by
\ba
\tilde{P}_{\rm fit}(k_i,\mu_p)&=&\left[P_{gg}(k_i) + 2\mu_p^2\sqrt{P_{gg}P_{\Theta\Theta}} + \mu_p^4P_{\Theta\Theta}(k_i)\right] \nn\\
&\times&G(k,\mu_j,\sigma_v)\,. \label{eq:fitpk}
\ea
We consider the velocity dispersion effect from one--dimensional velocity dispersion $\sigma_v$ which is given by,
\ba
\left(\frac{\sigma_v}{aH}\right)^2=\frac{1}{3\cdot 2\pi^2}\int
P_{\Theta\Theta}(k,z) dk\,. \label{eq:sigma_v}
\ea
This formula needs $P_{\Theta\Theta}$ which is what we want to
measure. We will discuss how we calculate this term in the next paragraph.
Eq.~\ref{eq:fitpk}
is expected to be invalidated beyond some
threshold. The observed modes are cut out when it goes beyond given the
threshold limit $\sigma_{\rm th}$ as $k_i\mu_{\rm cut}\sigma_v>\sigma_{\rm th}$. The fiducial value is $\sigma_{\rm th}=0.24$ which represents confidence of theoretical prediction up to 6$\%$ drop of $G(k_i,\mu_{\rm cut},\sigma_v)$ from unity.

The most important factor in the integration Eq.~\ref{eq:sigma_v}
is the amplitude of $P_{\Theta\Theta}$, as scale--dependent factor of $P_{\Theta\Theta}$ is
tightly constrained by CMB physics.
The shape of the power spectra is determined before the epoch of matter--radiation equality. When the initial fluctuations reach the coherent evolution epoch after matter-radiation equality, they experience a scale-dependent shift from the moment they re-enter the horizon to the equality epoch. Gravitational instability is governed by the interplay between radiative pressure resistance and gravitational infall. The different duration of modes during this period results in a shape dependence on the power spectrum. This shape dependence is determined by the ratio between matter and radiation energy densities and sets the location of the matter-radiation equality in the time coordinate~\citep{Song:2010vh}. 

One way to estimate $\sigma_v$ will be to use fitted
$P_{\Theta\Theta}$ for each fitting step. Our measurement is, however,
limited at scale of $k\lesssim 0.1\hompc$ and the contribution to
$\sigma_v$ from $P_{\Theta\Theta}$ at $k\gtrsim 0.1\hompc$ is small but
not negligible ($\sim 10\%$). Therefore, we calculate $\sigma_v$ using
the linear shape of $P_{\Theta\Theta}$ with an amplitude which is
estimated at each fitting step as follows.

For each $P_{\Theta\Theta}$ we want to test, we calculate the amplitude factor
$g_{\Theta}(k_i,z)$ defined by
\ba
P_{\Theta\Theta}(k_i,z)=g^2_{\Theta}(k_i,z)P_{\Theta\Theta}(k_i,z_{\rm lss}),
\ea
and constrain the amplitude by calculating a weighted average of
\begin{equation}
\bar{g}_{\Theta}(z)=\frac{\sum^{i_{max}}_{i=i_{min}}\left( g_{\Theta}(k_i,z)/\sigma^2_{g_{\Theta}}(k_i,z) \right)}{\sum^{i_{max}}_{i=i_{min}}1/\sigma^2_{g_{\Theta}}(k_i,z)}.
\end{equation}
Here $\sigma_{g_{\Theta}}(k_i,z)$ is given by
\ba
\sigma_{g_{\Theta}}(k_i,z)=g^{\rm fid}_{\Theta}(k_i,z)\frac{\sigma[P^{\rm fid}_{\Theta\Theta}(k_i,z)]}{P^{\rm fid}_{\Theta\Theta}(k_i,z)}\,,
\ea
and $\sigma[P^{\rm fid}_{\Theta\Theta}(k_i,z_j)]$ is given by theoretical estimation in Eq.~\ref{eq:sigP} and superscript `fid' denotes the fiducial model for Fisher matrix analysis.
We would not expect that fractional error of $P_{\Theta\Theta}(k_i,z)$ is much dependent on different fiducial models.
The value of $\sigma_v$ at the best fitted power spectra is
$2.8\mpcoh$  (the linear theory prediction is $3.2\mpcoh$).

$P_{gg}(k_i)$ determines the overall amplitude of $\tilde{P}_{\rm
fit}(k_i,\mu_j)$, and $P_{\Theta\Theta}(k_i)$ determines the running of
$\tilde{P}_{\rm fit}(k_i,\mu_j)$ in the $\mu$ direction. These distinct
contribution allows us to separate information of $P_{gg}(k_i)$ and
$P_{\Theta\Theta}(k_i)$ from 5 different $\mu$ bins at each $k_i$
bin. We find these $P_{gg}(k_i)$ and $P_{\Theta\Theta}(k_i)$ by minimizing
\ba
\chi^2&=&\sum^{i_{max}}_{i=i_{min}}\sum^{5}_{p=1} \sum^{5}_{q=1}  
[\tilde{P}_{\rm ob}(k_i,\mu_p)-\tilde{P}_{\rm fit}(k_i,\mu_p)]\nn\\
&\times&{\rm Cov}^{-1}_{pq}(k_i)
[\tilde{P}_{\rm ob}(k_i,\mu_q)-\tilde{P}_{\rm fit}(k_i,\mu_q)]\,,
\ea
where $k_{i_{min}}=0.03\hompc$ and $k_{i_{max}}=0.09\hompc$.
Off diagonal elements of the covariance matrix are nearly negligible and those diagonal elements are written as
\ba
{\rm Cov}^{-1}_{pp}(k_i)=\frac{1}{\sigma[\tilde{P}_{\rm ob}(k_i,\mu_p)]^2}\,.
\ea
We present the difference between $\tilde{P}_{\rm th}(k_i,\mu_p)$
(Kaiser effect) and $\tilde{P}_{\rm fit}(k_i,\mu_p)$ (including
dispersion effects) in Fig.~\ref{fig:Pks}. With the fiducial
$\sigma_{\rm th}=0.24$, only one bin of mode $k_i=0.09\hompc$ at
$\mu_p=0.9$ is removed from fitting. 
Altough this fitting procedure leads to correlations among different $k$
bins through $\sigma_v$, those are minimally correlated and the results shown Fig.~\ref{fig:contour_dP_dec} are consistent with theoretical predictions.

\section{Results and Discussion}

Velocity--velocity spectra are remarkably well extracted out of measured spectra in redshift space at scales $k=0.03, 0.05$, and $0.07\hompc$, and relatively well extracted at scale $k=0.09\hompc$ with more conservative confidence on the threshold limit. Filled blue contours in Fig.~\ref{fig:contour_dP_dec} represent fitted value of $P_{gg}(k_i)$ and $P_{\Theta\Theta}(k_i)$, and unfilled black contours represent estimation from theory with central values given by simulation. For scales from $k=0.03$ to $0.07\hompc$, the decomposed $P_{\Theta\Theta}(k_i)$ though our fitting strategy is trustable, which suggests that the few assumptions made in this paper are valid for those scales: 
\begin{itemize}
\item{The assumption of perfect correlation between halo distribution and velocity field is correct. 
The agreement of $P_{\Theta\Theta}(k_i)$ between fitted and true values supports our assumption of $r_h\sim 1$ indirectly.}
\item{Dispersion effect is reasonably modelled at scales within our confidence limits, which enables us to extract $P_{\Theta\Theta}(k_i)$ in model independent way using estimated $\sigma_v$.}
\end{itemize}

\begin{figure}
\begin{center}
\resizebox{3.2in}{!}{\includegraphics{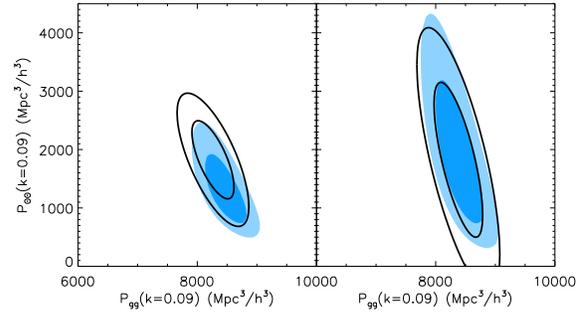}}
\end{center}
\caption{Contour plots are shown for decomposed $P_{gg}(k_i)$ and $P_{\Theta\Theta}(k_i)$ at $k=0.09\hompc$  with $\sigma_{\rm th}=0.24$ (left panel) and 0.18 (right panel). Unfilled black and filled blue contours represent the same in Fig.~\ref{fig:contour_dP_dec}.}
\label{fig:contour_Pk_mucut_dec}
\end{figure}

For $k=0.09\hompc$, more conservative threshold limits should be applied to remove non-linear supression. In Fig.~\ref{fig:contour_Pk_mucut_dec}, we present best fit $P_{\Theta\Theta}(k_i)$ with different threshold limits of $\sigma_{\rm th}=0.24$ (left panel) and $\sigma_{\rm th}=0.18$ (right panel). With $\sigma_{\rm th}=0.24$, only one bin at $\mu_j=0.9$ is removed. Shown in Fig.~\ref{fig:Pks}, extra suppression is also observed at $\mu_j=0.7$ bin at $k=0.09\hompc$ which can be removed by more conservative bound $\sigma_{\rm th}=0.18$. Shown in the right panel of Fig.~\ref{fig:contour_Pk_mucut_dec}, true $P_{\Theta\Theta}(k_i)$ is restored at the cost of weakened constraint.

Theoretical estimation from Fisher matrix analysis is an optimistic bound on errors. It is noticeable that measured varinaces (filled blue contours in Fig.~\ref{fig:contour_dP_dec}) are consistent with estimated variances (unfillled black contours in Fig.~\ref{fig:contour_dP_dec}), which assures us that our method is optimized extraction of $P_{\Theta\Theta}(k_i)$ for the given simulation specification.

\section{Conclusion}

We propose a statistical tool to decompose $P_{gg}(k)$ and $P_{\Theta\Theta}(k)$ practically out of redshift distortion maps, with a few assumptions: 1) perfect correlation between density and velocity fluctuations, 2) confidence on theoretical prediction of velocity dispersion effect within threshold limit. The results show that the true value of velocity--velocity spectra up to $k=0.07\hompc$ are successfully recovered using theoretical dispersion effect. The detectability is extendable up to $k\sim 0.1\hompc$ with more conservative threshold limit at the cost of weakened constraint. We find that the theoretical dispersion effect can be estimated from $P_{\Theta\Theta}(k)$ parameters using weighted average at $k<0.1\hompc$. In linear regime, $P_{\Theta\Theta}(k)$ is well--measured with this estimated $\sigma_v$ as much as with the true fixed $\sigma_v$ of the simulation. 

We find that the biased measurement of $P_{\Theta\Theta}(k)$ is mainly
caused by the unpredictable non--linear supression effect at $k>0.1\hompc$. The detectability limit in scale can
be extended by parameterizing this effect \citep{Tang10}, but we scope our range of interest in linear regime in this paper.

\section*{Acknowledgments}

We would like to thank Romain Teyssier for offering simulation map of
Horizon ({\it http://www.projet-horizon.fr}), and to thank Prina Patel for comment on the manuscript. Y-S.S. is supported by STFC and I.K. acknowledges support by
JSPS Research Fellowship and WPI Initiative, MEXT, Japan.

% \bibliographystyle{mn2e}
% \bibliography{ref}

\end{document}